\documentclass[twocolumn,iop,apj,twocolappendix]{aastex63}
\usepackage{amsfonts,amsmath,graphicx,natbib,url,hyperref,nicefrac}
\usepackage{amssymb, verbatim, subfigure, paralist, soul, comment}
\soulregister{\citep}{7}
\usepackage{float, tabularx, booktabs}
                    
\usepackage{verbatim}
\usepackage{tikz}
\usepackage{color}

\usepackage[T1]{fontenc}
\usepackage{aecompl}
\usepackage{calligra}
\DeclareMathAlphabet{\mathcalligra}{T1}{calligra}{m}{n}
\DeclareFontShape{T1}{calligra}{m}{n}{<->s*[2.2]callig15}{}

\usepackage{hyperref}
\hypersetup{
    pdfnewwindow=true,      % links in new window
    colorlinks=true,       % false: boxed links; true: colored links
    linkcolor=orange,          % color of internal links
    citecolor=cyan,        % color of links to bibliography
    filecolor=blue,      % color of file links
    urlcolor=blue           % color of external links
}

\def\apjl{ApJL}               % Astrophysical Journal, Letters
\def\apjs{ApJS}               % Astrophysical Journal, Supplement
\def\aap{A\&A}                % Astronomy and Astrophysics
\DeclareMathAlphabet{\mathcalligra}{T1}{calligra}{m}{n}
\DeclareFontShape{T1}{calligra}{m}{n}{<->s*[2.2]callig15}{}

\usepackage{color}

\usepackage{comment}

\shorttitle{Circumbinary Disk Rarity} 
\shortauthors{Balivada and Duffell}

\begin{document}

\title[]{On the Evolution of Circumstellar Disks Modified by the Presence of a Binary}

\author[0000-0002-6154-4170]{Aishwarya Balivada}
\affiliation{Department of Physics and Astronomy, Purdue University, 525 Northwestern Avenue, West Lafayette, IN 47907, USA}
%\email{abalivad@purdue.edu}

\author[0000-0001-7626-9629]{Paul C. Duffell}
\affiliation{Department of Physics and Astronomy, Purdue University, 525 Northwestern Avenue, West Lafayette, IN 47907, USA}
%\email{pduffell@purdue.edu}

\begin{abstract}\label{sec:abstract}

Stars form within circumstellar disks, and some binary stars develop circumbinary disks surrounding the pair. Observations suggest that disks that form around binary star systems are less common than those around single stars. One possible explanation is that binary stars may disrupt or shorten the lifespan of their surrounding disks relative to those around single stars. To investigate this, we used the DISCO hydrodynamics code to compute the lifetimes of circumstellar disks with and without a binary. Our results indicate that binary systems can reduce the lifetime of a circumbinary disk by a modest factor of two to three compared to disks around single stars. This effect is present when the binary separation is at least one-tenth of the disk radius. While our findings show that binaries can shorten circumbinary disk lifetimes, this alone might not be enough to fully account for the scarcity of these disks.
\end{abstract}

\section{Introduction}\label{sec:introduction}

\begin{figure}
\centering
\includegraphics[width=3.2in]{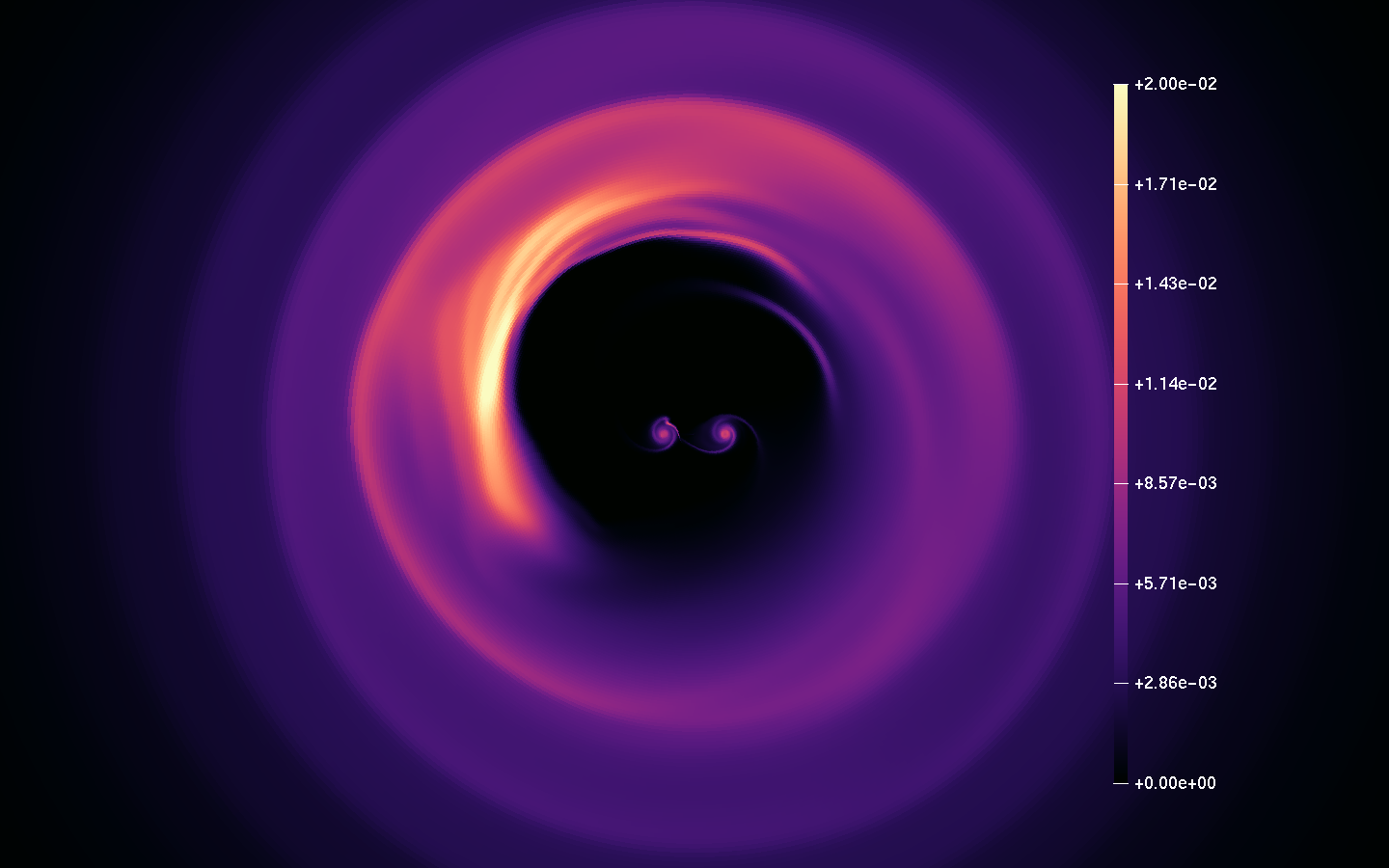}
\caption{Surface density profile of a circumbinary disk with R$_{d} = 2a$. Binary-induced gravitational perturbations are evident. The spatial average of this surface density as a function of time will be used to track the disk evolution.} 
\label{fig:fig_1}
\end{figure}

After the collapse of molecular clouds, stars typically form within disks of gas and dust, with the majority of these stars possibly forming in binary systems (\cite{1991A&AS...88..281D}; \cite{1991A&A...248..485D}; \cite{2023A&A...674A...1G}). The demographics of binary systems are consistent with the idea that most binaries form out of the same molecular cloud (see also \cite{2024MNRAS.532.2374M}). In such systems, the process of star formation is accompanied by the development of circumstellar disks. Each star in the binary system may be surrounded by its own individual disk: the more massive primary star can host a circumprimary disk, while the less massive secondary star can host a circumsecondary disk. In addition to these individual disks, binary systems can also host a circumbinary disk that orbits both stars and their disks, encompassing the entire system (See Figure \ref{fig:fig_1}).  All three types of disks have been observed in binary T Tauri systems such as GG Tau (\cite{1999A&A...348..570G}), DQ Tau (\cite{2024AJ....168..168Y}), UZ Tau E (\cite{2020ApJ...900....7H}), HD 142527 (\cite{2010PASJ...62..347F}), and HD 100453 \cite{2010PASJ...62..347F}).

Observations of young stars have compared the prevalence of circumprimary, circumsecondary, and circumbinary disks to disks around single stars (\cite{2000prpl.conf..703M}; \cite{2012ApJ...751..115H}). Studies have found that circumprimary/circumsecondary disks around each component of wide separation binary systems are more common than circumbinary disks, which are typically found in systems with smaller separations, seen, for example, in Figure $6$ of \cite{2012ApJ...751..115H}. These observations suggest that circumbinary disks are rarer than circumprimary/circumsecondary disks, raising the question of whether the presence of a binary system inhibits disk formation or enhances disk dissipation. 

This distinction is worth emphasizing, since approximately half of all solar-type stars are in binary or multiple-star systems (\cite{2010ApJS..190....1R}; \cite{2012ApJ...745...19K}). Observational evidence shows that close binary companions can reduce the lifetime of circumstellar disks compared to those around single stars (\cite{2009ApJ...696L..84C}).
In close binary systems, gravitational interactions between the stars and the surrounding disk can truncate the disk’s outer edge and excite spiral density waves that redistribute angular momentum. These effects restrict the disk’s radial extent and increase the efficiency of angular momentum transport, both of which shorten the disk’s overall lifetime (\cite{1996ApJ...467L..77A}; \cite{2005ApJ...623..922O}; \cite{2014prpl.conf..475A}; \cite{2011ARA&A..49..195A}). 

This raises a key question: are circumbinary disks intrinsically rare, or do they simply dissipate more quickly in the presence of a binary? In this work, we carry out a fundamental study of how binaries affect disk evolution at high viscosity values, using scaling arguments to extend our findings toward lower viscosities thought to be more realistic in disks (see, e.g., \citep{2012ApJ...751..115H}). Essentially, we are treating these models as an idealized set up to identify the structural and accretion response to the binary. For simplicity, we adopt a spatially uniform kinematic viscosity $\nu$. This assumption is an idealization used to clarify the influence of the binary. Because the models are initialized with a finite mass reservoir, the surface density evolves with time and the accretion rate gradually declines once truncation occurs. Although these viscosity values are intentionally higher than those inferred from recent observations, they are adopted here as an idealized choice to highlight how circumbinary disk lifetimes scale with binary separation and viscous transport. For comparison, the DSHARP survey \citep{2018ApJ...869L..46D} finds effective $\alpha$-values two orders of magnitude lower than the values used here, and 3D circumbinary simulations such as \citep{2021MNRAS.508.4806P} similarly report reduced levels of turbulence generated by the binary. Our goal in this first study is therefore not to reproduce physical turbulence, but to use a controlled and computationally efficient parameter range to establish how lifetime trends depend on viscosity in an idealized framework.
 
While it is clear that binaries influence disk evolution, it remains uncertain how binary separation and disk viscosity together shape circumbinary disk lifetimes. In this work, we carry out a fundamental study of how binaries affect disk evolution at modifiable viscosity values, using scaling parameters to extend our findings toward the lower viscosities thought to be more realistic in disks (see, e.g., \cite{2014prpl.conf..475A}).

The dynamics of circumbinary disks have been studied extensively (\cite{2005ApJ...623..922O}; \cite{2017MNRAS.466.1170M}; \cite{2019ApJ...871...84M}; \cite{2019ApJ...875...66M}; \cite{2020ApJ...901...25D}; \cite{2020A&A...641A..64H}; \cite{2026arXiv260627419R}; for review see \cite{2023ARA&A..61..517L}). Previous reviews, such as those by \cite{2000prpl.conf..731L}, have explored how the gravitational interactions between the stars and the disk influence the structure and evolution, including the formation of circumstellar and circumbinary disks and the effects of binary system dynamics on disk gaps and resonances. Disks may be truncated at a specific disk size due to the binary's gravitational influence. Positive torques generated by these interactions can drive material away from the inner region of the disk (\cite{2000prpl.conf..731L}), impacting both its accretion rates and the disk lifetime. 

We investigate the possibility that circumbinary disks are less common because their lifetimes are shorter. To do this, we use the DISCO hydrodynamical code (\cite{2016ApJS..226....2D}) to simulate the evolution of circumbinary disks with different disk sizes (R$_{d}$), binary separation ($a$), and viscosities ($\nu$). These simulations are carried out in both one-dimensional (1D) and two-dimensional (2D) configurations to examine how the disk lifetime $\tau$ changes under each setup. In the 1D model, the binary is replaced by a single star. In contrast, the 2D model includes the full binary system, allowing us to capture the complex interactions between the stars and the disk, including resonances, gaps, and changes to the disk's structure. By comparing these two models, we aim to determine whether the binary stars shorten the lifetime of the circumbinary disk.

The remainder of this paper is organized as follows. In \ref{sec:methods}, we describe how DISCO is used to simulate the dynamics of circumbinary disks. Section \ref{sec:results} compares disk lifetimes at different viscosities in single-star and binary star systems obtaining a disk suppression factor. Section \ref{sec:discussion} introduces an analytical expression that summarizes all the data we find empirically. 

\section{Methods}\label{sec:methods}

To investigate the dynamics of circumbinary disks, we employed the DISCO code, a numerical hydrodynamics solver that models viscous mass transport in disks \citep{2016ApJS..226....2D}. Using this code, we implemented an equal mass binary system in a circular Keplerian orbit surrounded by a finite disk with a disk size, R$_{d}$, and initialized the disk based on the initial model used in the Santa Barbara Code Comparison \citep{2024ApJ...970..156D}. That study used the following initial condition for the surface density.

\begin{equation}
    \Sigma(r) = \Sigma_0[(1 - \delta_0)e^{-(R_{cav}/r)^{12}} + \delta_0] f(r)
\label{eq:eq_sigmar}
\end{equation}
where $R_{cav} = 2.5 a$ and $\delta_{0} = 10^{-5}$. Here $\Sigma_0$ is a scaling parameter representing the surface density at the reference radius (typically near the inner edge of the disk), not the initial density across the entire disk. The term $(1 - \delta_{0}) e^{-(R_{\text{cav}}/r)^{12}}$ models the distribution of surface density in the disk, with $\delta_{0}$ controlling a small background density. The parameter $R_{\text{cav}}$ denotes the radius of a cavity in the disk created by the binary's gravitational influence, while $f(r)$ is taken to be an exponential function of radius: 

\begin{equation}
    f(r) = e^{-r/R_d}.
\label{eq:eq_cavityrad}
\end{equation}
Here, R$_{d}$ represents the characteristic disk size. Disk sizes ranged from R$_{d} \sim a$ to R$_{d} = 20 a$, with varying viscosity values. These models were run in two modes: the standard binary setup was run in two dimensions (2D), but followed by a one-dimensional (1D) calculation that replaced the binary with a single-point mass. With both modes we are able to measure how the presence of the binary for each case affects the disk lifetime.

The viscosity is specified as a constant value $\nu$ throughout the disk. Although uniform $\nu$ would correspond to a steady-state solution only if the surface density were also constant, our disks are initialized with finite mass and therefore do not approach such a state. Instead, this treatment provides a simplified framework for isolating the combined viscous and tidal effects on disk evolution. 

\subsection{Boundary Conditions} \label{subsec:boundary conditions}

The computational domain extends to an outer radius of at least twice the initial disk size, $R_{\mathrm{out}} \gtrsim 2R_d$, so that the boundary does not influence the disk evolution over the model time span. At the inner edge, we employ the standard sink cell treatment implemented in DISCO (\citep{2016ApJS..226....2D}, \citep{2024ApJ...970..156D}), which fixes the density and velocity to values that effectively remove material once it crosses within one binary separation. This prescription mimics accretion onto the stars and prevents artificial accumulation of gas.

\subsection{Influence of Initial Transients}

\begin{figure}
    \centering
    \includegraphics[width=0.45\textwidth]{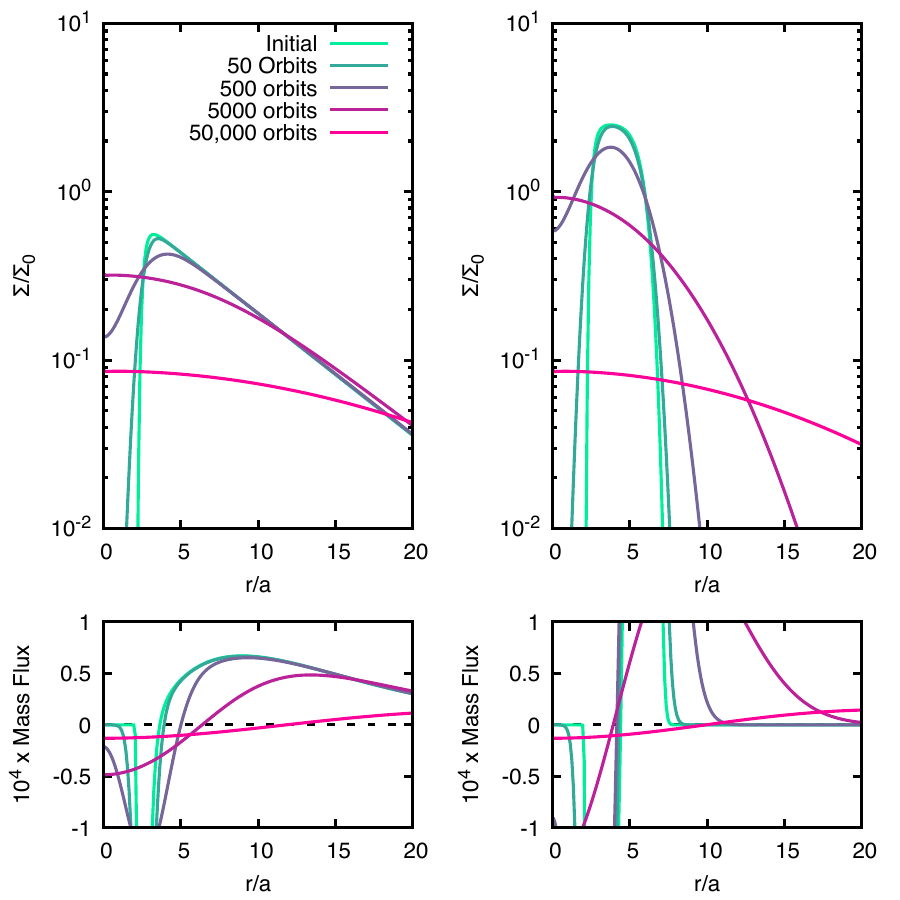}
    \caption{Evolution of the one-dimensional disk models from two different initial surface-density profiles. The upper panels show the surface density normalized by $\Sigma_{0}$, while the lower panels show the corresponding radial mass flux. Profiles are shown for the initial condition and after 50, 500, 5,000, and 50,000 binary orbits. Although the models differ during their early adjustment, they approach the same long-term viscous evolution. The corresponding initial disk masses are $206.8\,\Sigma_0 a^2$ and $209.7\,\Sigma_0 a^2$ for the sharp-cutoff and exponential profiles, respectively.}
    \label{fig:initial_profiles}
\end{figure}

To assess the role of the initial disk configuration, we compared our exponential surface-density profile with a second profile in which the surface density decreases more sharply with radius. The early-time adjustment phase lasts for the first few hundred binary orbits, during which both setups relax from their initial conditions. After this settling period, the two disks follow the same long-term depletion trend, indicating that the lifetime measurements reported in this work do not depend on the particular initial surface-density profile chosen. This comparison is shown in Figure~\ref{fig:initial_profiles}. For the sharp-cutoff and exponential initial profiles, the initial disk masses are $M_{\rm d}=206.8\,\Sigma_0 a^2$ and $M_{\rm d}=209.7\,\Sigma_0 a^2$, respectively.

\section{Results}\label{sec:results}

Following the modeling guidelines in \cite{2024ApJ...970..156D}, we performed high-resolution simulations up to $5000$ orbits, as shown in Figure \ref{fig:fig_1}. Figure \ref{fig:fig_1} shows the surface density profile and accretion flow in a 2D simulation with R$_{d} = 2a$. Over time, this surface density will decrease due to accretion onto the binary and viscous spreading of the disk.

\begin{figure}
\includegraphics[width=\columnwidth, height=2.5in, width=3.5in]{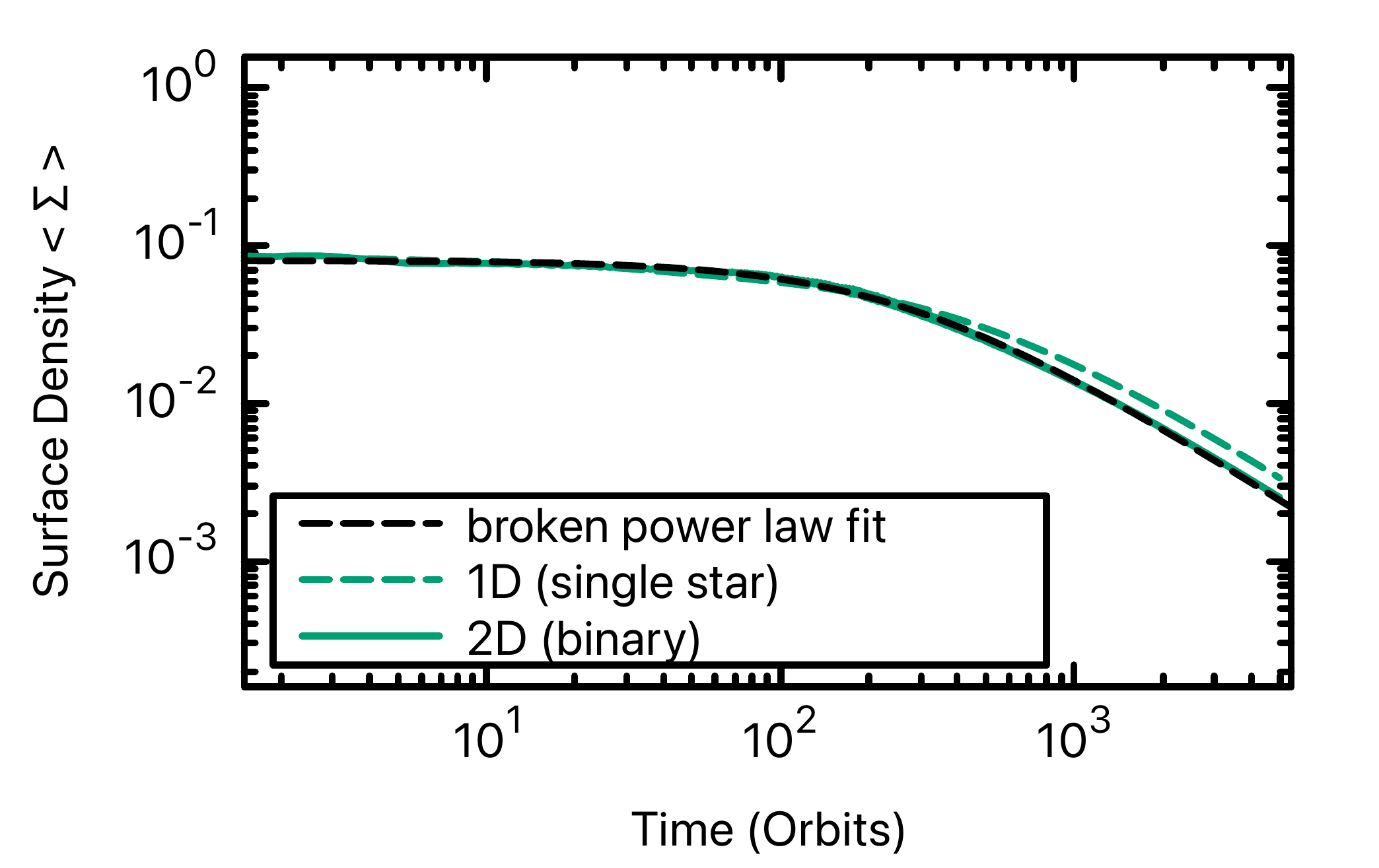}
\caption{Average surface density over time for $5000$ orbits, with disk radius $R_{d} = 2a$ and viscosity $\nu = 10^{-3} a^2 \Omega_B$. Comparison is made between 1D (green dashed line) and 2D (green solid line) DISCO models. The 2D model has a shorter lifetime than the 1D model due to the tidal interaction from the binary in 2D. The black dashed line represents a fitting function (Equation \ref{eq:eq_ARdi}) used to extract the disk lifetime.} 
\label{fig:fig_2}
\end{figure}

Figure \ref{fig:fig_2} shows the average surface density in the disk over time. To compute surface density, we use: 

\begin{equation}
    \Sigma_1 = \int \Sigma^2 dV, \quad \Sigma_2 = \int \Sigma dV.
\label{eq:eq_sigma1_and2}
\end{equation}

The surface density is then given by the ratio:

\begin{equation}
    \left< \Sigma \right> = \frac{\Sigma_1}{\Sigma_2} 
    = \frac{\int \Sigma^2 dV}{\int \Sigma dV} 
    = \frac{\int \Sigma dm}{\int dm}. 
\label{eq:eq_Sigma}
\end{equation}
As seen in Figure \ref{fig:fig_2}, the time dependence $\Sigma(t)$ exhibits a broken power law behavior with a distinct drop after it has reached a characteristic $\tau$ value for R$_{d} = 2a$. To find the disk lifetime $\tau$, we fit $\Sigma(t)$ using a broken power law:
\begin{equation}
    \Sigma(t)= \frac{A}{(1 + (\frac{t}{\tau})^k)t^n},
\label{eq:eq_ARdi}
\end{equation}
where $A$, $k$, $\tau$, and $n$ are fitting parameters.

\begin{figure}
\includegraphics[width=\columnwidth, height=2.5in, width=3.5in]{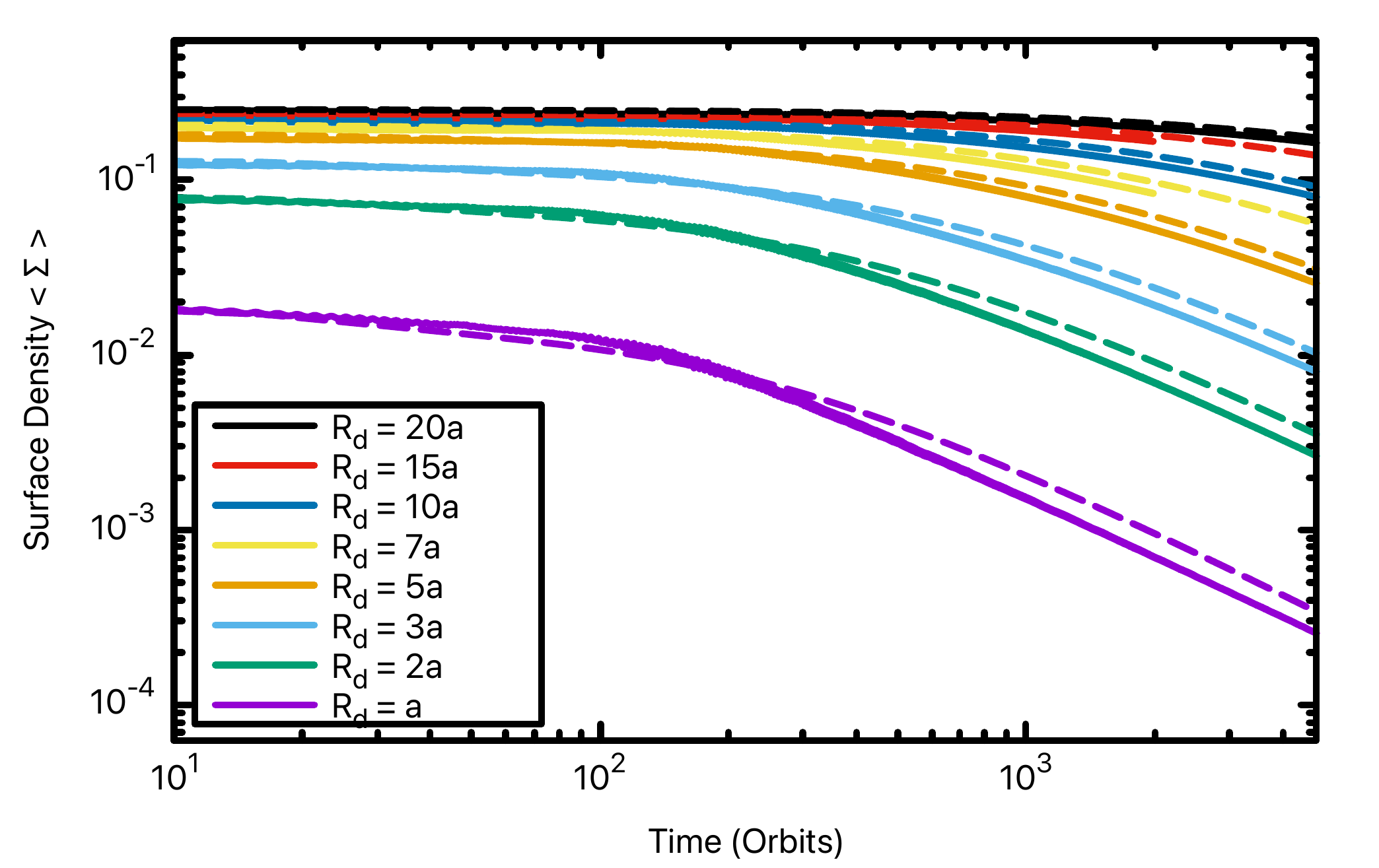}
\caption{Average surface density over time for a range of different disk radii ($R_{d}$), showing a consistent discrepancy between 1D and 2D disk lifetime. Comparison is made between 1D (dashed line) and 2D (solid line) DISCO models.} 
\label{fig:fig_3}
\end{figure}

Figure \ref{fig:fig_3} resembles Figure \ref{fig:fig_2}, but presents the average surface density for various disk sizes over time. The results indicate that larger disks tend to survive for longer lifetimes. An examination of the radial mass flow shows the expected pattern: gas moves inward in the inner disk due to binary torques, while the outer disk spreads outward more slowly under viscosity. The radius at which the radial mass flux changes sign moves outward as the disk evolves. For the disk sizes studied here, overall depletion is dominated by accretion onto the binary rather than outward spreading, consistent with the shorter lifetimes observed in 2D calculations.

In absence of the binary, one can estimate the disk lifetime based off of the viscous timescale:
\begin{equation}
    \tau \sim \frac{R_{d}^2}{\nu}.
\label{eq:eq_tau}
\end{equation}

\begin{figure}
\includegraphics[width=\columnwidth, height=2.5in, width=3.5in]{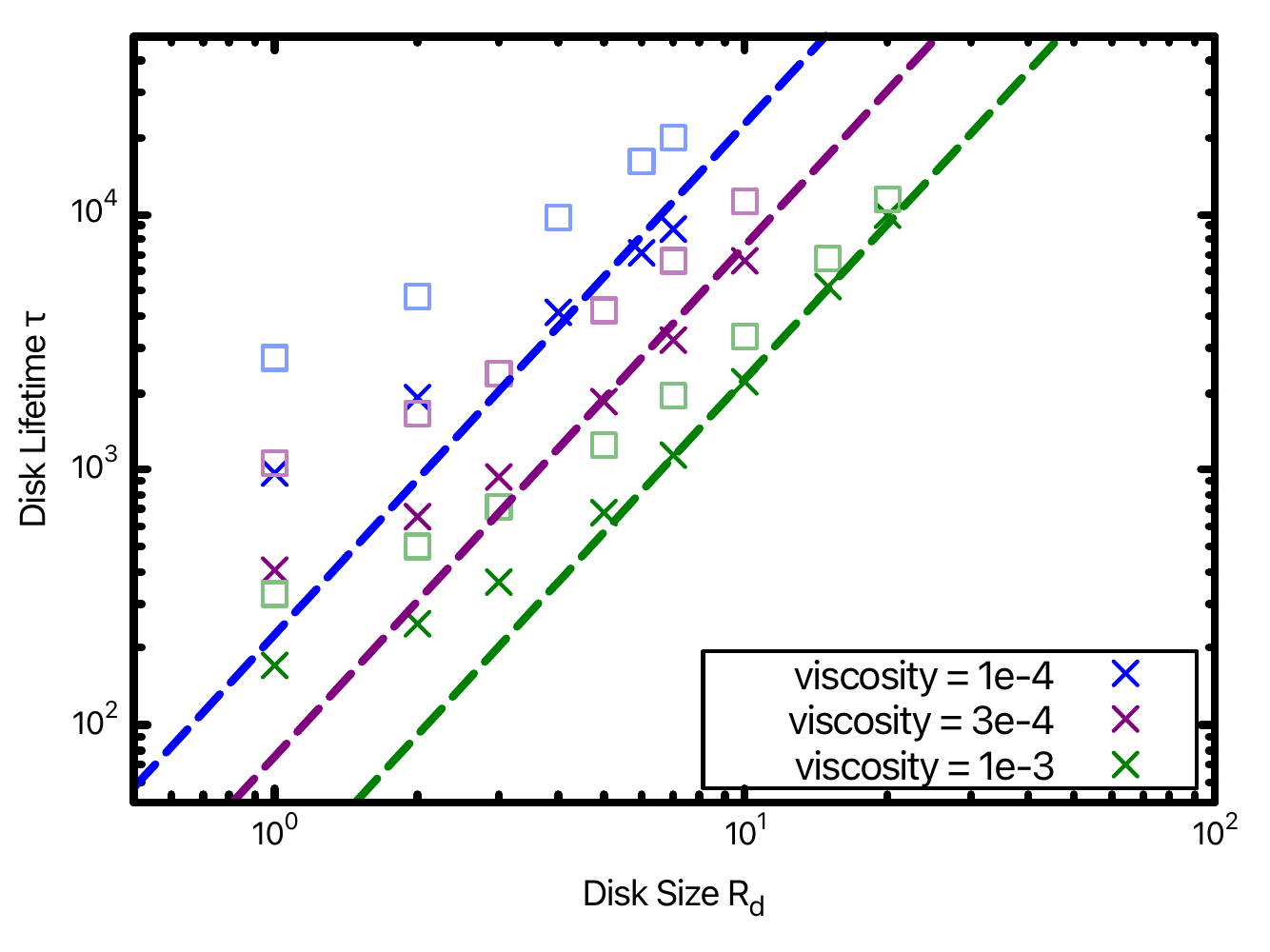}
\caption{Disk lifetime ($\tau$) as a function of disk radius ($R_d$) comparing 1D (squares) and 2D ('X') DISCO simulations. The dashed curve represents an analytical estimate for the lifetime (Equation \ref{eq:eq_tau}). For disks where $R_{d} > 10a$, the 1D (single-star) and 2D (binary) results converge to each other and to the analytical estimate (dashed line).
Here $\tau$ is expressed in binary orbits, and $R_{d}$ is given in units of the binary separation $a$.}
\label{fig:fig_4}
\end{figure} 

Figure \ref{fig:fig_4} highlights $\tau$ as a function of R$_{d}$, showing convergence of 1D and 2D results with analytical predictions for R$_{d} > 10a$. For large R$_{d}$, the 1D and 2D results both converge to the analytical estimate. For smaller R$_{d} \sim a$, the systems with a binary have shorter lifetimes than the same disk sizes around a single star. The ratio of $\tau_{\rm 1D} / \tau_{\rm 2D}$ gives a disk suppression factor due to the influence of the binary. We compute single-to-binary lifetime ratios ranging from $2$ to $2.8$ for each disk, meaning that the binary reduces the disk lifetime by this factor. These results are summarized in Figure \ref{fig:fig_5}.

\section{Discussion}\label{sec:discussion}

This study examines the dynamics and lifetimes of circumbinary disks compared to circumstellar disks around single stars. In the following, we summarize the primary results, their implications, and the role these findings play in understanding the rarity of circumbinary disks.

\begin{figure}
\includegraphics[width=\columnwidth, height=2.3in, width=3.5in]{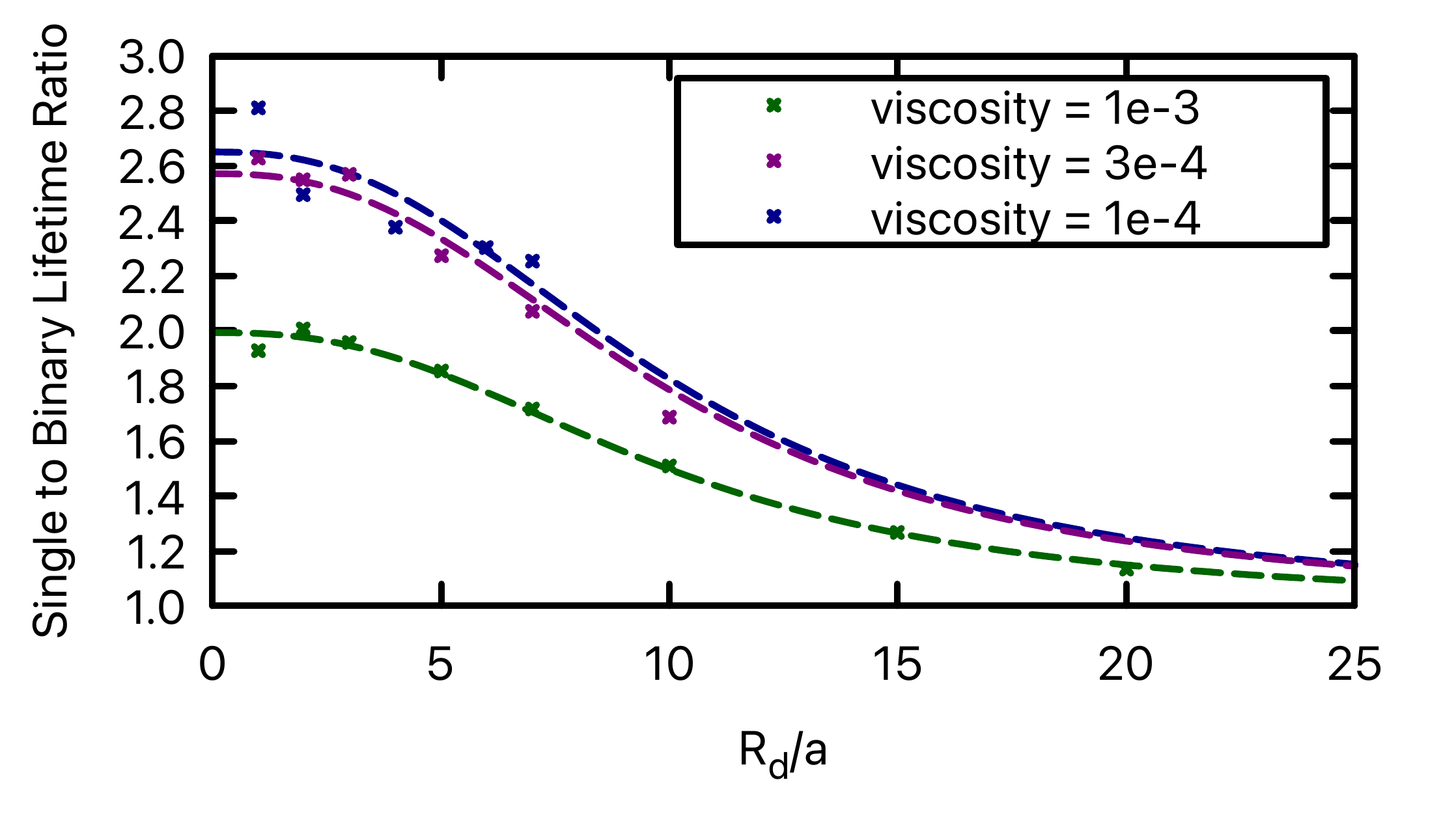}
\caption{Ratio of disk lifetime ($\tau$) between single-star and binary systems, $\tau_{\rm 1D} / \tau_{\rm 2D}$, as a function of the disk-to-binary separation ratio ($R_{d}/a$). In our models, binary systems reduce the lifetime of a circumbinary disk $(\tau_{\rm 2D})$ by a modest factor of two to three compared to disks around single stars $(\tau_{\rm 1D})$. This effect is present when the disk size $R_d$ is less than ten times the separation $a$. Equivalently, the disk lifetime is shortened by the binary when the binary separation is at least one-tenth of the disk radius.} 
\label{fig:fig_5}
\end{figure}

Our results show that circumbinary disks typically have lifetimes two to three times shorter than circumstellar disks around single stars. This effect is most pronounced when the binary separation $a$ is at least one-tenth the radius of the disk, R$_{d}$ (Figure \ref{fig:fig_5}). For larger disks where R$_{d} \gg a$, the lifetime scales as $\tau \propto R_{d}^2$, with minimal dependence on binary separation. When R$_{d} \approx a$, binary interactions dominate, significantly shortening $\tau$.

In order to summarize all the data, we found through trial-and-error that the results are reasonably well-fit to the following analytical expression:
\begin{equation}
    \frac{\tau_{\rm 1D}}{\tau_{\rm 2D}}  = \frac{\beta - 1}{(1 + \frac{R_d}{10a} )^{2.5}} + 1.
\label{eq:eq_ARdf}
\end{equation}
where $\beta$ is the lifetime ratio in the limit of small disk size ($R_d \rightarrow 0$). In general, $\beta$ depends on the disk viscosity but for our disk models we find this factor is consistently around 2-3 (for $\nu = 10^{-3}a^2\Omega_B$ we find $\beta = 2.1$, for $3 \times 10^{-4}a^2\Omega_B$ we find $\beta = 2.62$ and for $10^{-4}a^2\Omega_B$ we find $\beta = 2.7$).

Our results indicate that binary systems can reduce the lifetime of a circumbinary disk by a modest factor compared to disks around single stars. Because of the factor ($R_d/10a$) in Equation (\ref{eq:eq_ARdf}), we find this lifetime-shortening effect due to the binary is present when the binary separation is at least one-tenth of the disk radius. While our findings show that binaries can shorten circumbinary disk lifetimes, the modest reduction of the lifetime by a factor of only 2-3 demonstrates that this alone might not be sufficient to fully account for the scarcity of these disks. This suggests that alternate causes should be considered for depleting the circumbinary disk, such as dynamical capture at high eccentricities of systems formed via cold core fragmentation, as seen in simulations such as \cite{2018MNRAS.475.5618B, 2023A&A...674A.196K}.

Although our parameter exploration includes only three viscosity values, the scaling relation $\tau \propto R^{2}/\nu$ provides a simple framework for estimating how the measured trends may shift toward lower viscosities. This extrapolation should be interpreted cautiously, since lower-viscosity discs require longer integrations and may introduce additional structural and numerical effects that are not captured by the present idealized survey.

One caveat of this work is that we model equal-mass, circular binaries with isothermal disks. Real systems often involve unequal masses or eccentric orbits. These setups don't account for eccentric or unequal-mass binaries, which would lead to stronger, asymmetric truncation and more variable accretion onto the stars. We also ignore self-gravity and radiative transfer--factors that might not matter early on, but could change the disk structure or slow down its dissipation over longer timescales. So the exact suppression factors we report here might shift if those effects are included.

Real circumbinary disks may exhibit nonlinear dependencies between viscosity, scale height, and temperature structure, as seen in \citep{2022MNRAS.513.6158D}, \citep{2020MNRAS.498.2936H}, and recent studies of circumbinary disk structure and thermodynamics \citep{2024MNRAS.532.3166P, 2025MNRAS.544.3312P}. These effects are not captured in the locally isothermal, uniform-$\nu$ framework adopted here. Our goal in this study is to isolate the global depletion response of the disk to binary torques, while recognizing that more detailed thermodynamic treatments could modify the disk morphology.

In addition, we have not extended our analysis to a full 2D characterization of the flow. Detailed studies of the circumbinary flow field, including spiral shocks and gas streams, have been presented in prior work on circular binaries (see the review of \citep{2023ARA&A..61..517L} and references they are in). Here we instead focused on a different question: quantifying how the presence of the binary shortens the disk lifetime. Our goal was to establish a clear scaling relation for this effect, rather than to reproduce the detailed morphology of the flow. As such, the suppression factors reported here should be interpreted as global measure of disk evolution, complementary to the flow-focused investigations in the literature.

Another caveat is that we have not examined precession-driven changes in the circumbinary disk. At higher viscosities, such effects can propagate farther into the disk and influence local gas velocities, but in our models the suppression factors show only a weak dependence on viscosity. The weak sensitivity was somewhat unexpected, and it suggests that while precession could shift the detailed velocity structure, the overall reduced lifetime we measure is robust against the simplification. 

Our use of a uniform viscosity is an intentional simplification and does not yield a steady-state accretion flow. Because the surface density is not fixed, the mass flux is not constant with radius, as illustrated in Figure~ \ref{fig:initial_profiles}. This behavior reflects the interplay between viscous spreading and binary torques.

Another source of uncertainty is the boundary conditions.
The inner boundary uses a sink cell to model accretion onto the binary, while the outer boundary is placed far beyond the disk edge ($R_{\mathrm{out}} \gtrsim 2R_d$) so that it does not feed back on the disk during the simulations. These assumptions are consistent with prior circumbinary disk calculations \citep{2024ApJ...970..156D}, but they may affect the long-term evolution. In particular, resolving circumprimary or circumsecondary disks would introduce additional viscous timescales and overflow channels not captured in the present setup. As a result, the disk lifetimes we report should be regarded as global trends, rather than predictions of detailed binary–disk mass exchange.

\acknowledgements

This research was supported in part by the National Science Foundation under grant AAG-2206299. Hydrodynamic calculations were performed on the \texttt{Petunia} computing cluster, located at Purdue University. We would like to thank and appreciate the anonymous referee for their very thorough and detailed review of our work.

\bibliography{main}\label{sec:references}

\end{document}